# Relationship of peak fluxes of solar radio bursts and X-ray class of solar flares: Application to early great solar flares

Keitarou MATSUMOTO ,[1,*] Satoshi MASUDA,[1] Masumi SHIMOJO ,[2,3] and Hisashi HAYAKAWA [1,4,5]

[1] Institute for Space–Earth Environmental Research, Nagoya University, Furo-cho, Chikusa-ku, Nagoya, Aichi 464-8601, Japan
[2] National Astronomical Observatory of Japan, National Institutes of Natural Sciences, 2-21-1 Osawa, Mitaka, Tokyo 181-8588, Japan
[3] The Graduate University for Advanced Studies, SOKENDAI, 2-21-1 Osawa, Mitaka, Tokyo 181-8588, Japan
[4] Institute for Advanced Research, Nagoya University, Furo-cho, Chikusa-ku, Nagoya, Aichi 464-8601, Japan
[5] Space Physics and Operations Division, RAL Space, Science and Technology Facilities Council, Rutherford Appleton Laboratory, Harwell Oxford, Didcot, Oxfordshire OX11 0QX, UK

*E-mail: keitaromatsumoto@isee.nagoya-u.ac.jp; km876@njit.edu

## Abstract

Large solar flares occasionally trigger significant space-weather disturbances that affect the technological infrastructures of modern civilization, and therefore require further investigation. Although these solar flares have been monitored by satellite observations since the 1970s, large solar flares occur only infrequently and restrict systematic statistical research owing to data limitations. However, Toyokawa Observatory has operated solar radio observations at low frequencies (at 3.75 and 9.4 GHz) since 1951 and captured the early great flares as solar radio bursts. To estimate the magnitudes of flares that occurred before the start of solar X-ray (SXR) observations with the Geostationary Operational Environmental Satellite (GOES) satellites, we show the relationship between microwave fluxes at 3.75 and 9.4 GHz and X-ray fluxes of flares that occurred after 1988. In total, we explored 341 solar flares observed with the Nobeyama Radio Polarimeters and Toyokawa Observatory from 1988–2014 and compared them with the SXR observations recorded by the GOES satellites. The correlation coefficient was approximately 0.7. Therefore, the GOES X-ray class can be estimated from the peak flux at 3.75 and 9.4 GHz with a large variance and an error of factor of 3 ($1\sigma$). Thus, for the first time, we quantitatively estimated the light curves of two early solar flares observed in 1956 February by the Toyokawa solar radio observations using the relationship between SXR thermal radiation and microwave nonthermal radiation (Neupert, 1968, ApJ, 153, 59).

# 1 Introduction

Solar flares are explosions in the solar corona that occasionally direct geo-effective coronal mass ejections and considerably influence the space weather as indicated by geomagnetic storms and ground level enhancement (GLE) (Gonzalez et al. 1994; Aschwanden 2012; Shea & Smart 2012; Usoskin 2017; Cliver et al. 2022). The solar flare in 2003 October posed tremendous impacts on our civilization (Doherty et al. 2004; Ivanov & Kharshiladze 2007). The magnitude of the flares was measured by the Geostationary Operational Environmental Satellite (GOES) in terms of the maximum flux of the soft X-ray (SXR) (Fletcher et al. 2011). In general, various physical properties such as temperature and volume emission measure can be obtained from the SXR observations of GOES. This is useful in understanding the behavior of a hot plasma created by a solar flare. From the point of view of space weather, a statistical study (Yashiro & Gopalswamy 2009) focusing on the relationship between GOES SXR observations and coronal mass ejections (CMEs) has also been conducted. Thus, if we estimate the GOES SXR class of past large flares, we can infer more information about them.

In most of the recent flare studies, GOES-class was used as a standard index of the magnitude of a solar flare. Prior to the operation of the GOES satellite, the magnitude of the flares was estimated based on proxy observations (Curto et al. 2016; Hayakawa et al. 2022a). Since the occurrence rate of large flares is low, the SXR estimation of these past flares is also important from a statistical prediction point of view and improves the statistical prediction of the largest solar flares. Historical evidence indicates the occurrence of a great solar flare in 1956 February that caused the greatest GLE in observational history (Notsuki et al. 1956; Rishbeth et al. 2009; Cliver et al. 2020; Usoskin et al. 2020). Prior to that, the earliest flare was observed in 1859 September and is linked with one of the strongest geomagnetic storms in modern history (Cliver & Dietrich 2013; Hayakawa et al. 2019, 2022b; Hudson 2021).

Certain studies have statistically investigated the flare characteristics (limitation and frequency of flares) (Elvidge & Angling 2018; Tsiftsi & Luz 2018). As reported, the peak flux of microwave radiation is strongly related to that of SXR (Hudson & Ohki 1972; Kawate et al. 2011). If this is the case, the magnitudes of pre-GOES flares can be estimated based on microwave data and analyses of the correlation between the microwaves and SXR within the coverage of the GOES satellite. These correlations enable us to explore case studies for quantitatively estimating the flare magnitudes of the earliest solar flare based on Toyokawa solar radio observations, as flares were earlier observed in terms of microwaves (3.75 and 9.4 GHz). Furthermore, information on large flares can be ascertained from the Toyokawa Observatory observations, as recorded in Tanaka and Kakinuma (1956).

Neupert (1968) derived the relationship between the nonthermal microwave flux $\Phi_{\mathrm{Micro}}$ and thermal SXR flux $\Phi_{\mathrm{SXR}}$ as follows:

$$\Phi_{\mathrm{SXR}} \propto \int_{t_0}^{t} \Phi_{\mathrm{Micro}} dt, \tag{1}$$

where $t_0$ denotes the initiation of the microwave burst. The relationship between SXR and hard X-ray (HXR) has been referred to as the Neupert effect (Hudson 1991) and examined appropriately (Dennis & Zarro 1993; Veronig et al. 2005). To date, quantitative estimation of the light curve of the thermal radiation from nonthermal radiation has remained challenging. Thereafter, the light intensity curve of SXR has been quantitatively drawn by normalizing with the estimated flare class values. These analyses allow us to facilitate the reconstruction of the SXR light curves of the early solar flares observed in 1956 based on the historical microwave records and the Neupert effect. If we can reconstruct the SXR light curve as well as the peak flux, we can understand the characteristics of an individual flare well. For example, if we know the duration of a flare, we can determine whether it is an LDE (long duration event) flare or an impulsive one, and we can guess that the flare is associated with a CME if it is an LDE flare (Yashiro & Gopalswamy 2009). It is also known that the time variation of the SXR light curve is related to the length of the loop and the duration of magnetic reconnection (Reep & Toriumi 2017). We could further investigate the characteristics of the flare from the information about the time variation.

The source data used in this study are profiled in section 2. Thereafter, the analysis results derived in this study are discussed in section 3. In section 4, we digitized the solar radio bursts reported by Tanaka and Kakinuma (1956) and derive their SXR light curves based on the study results.

# 2 Observations and data selection

## 2.1 Radio polarimeters at Nobeyama and Toyokawa

Nobeyama Radio Polarimeters (NoRP) have been in operation since 1994 to observe microwaves across a wide range (Nakajima et al. 1985; Shimojo & Iwai 2022) of frequencies: 1, 2, 3.75, 9.4, 17, 35, and 80 GHz. Similarly, Toyokawa Radio Polarimeters (ToRP) (Tanaka & Kakinuma 1957; Shimojo & Iwai 2022) have been in operation since 1957 to observe microwaves across a wide range of frequencies: 1, 2, 3.75, and 9.4 GHz. We can estimate the magnitudes of the earliest flares in Toyokawa solar radio observations using the data at 3.75 and 9.4 GHz. We analyzed the data of 17 GHz because we need to classify the types of microwave radiation in terms of optical depth. We referred to the Nobeyama Radio Polarimeters Event List[1] to collect microwave data.

## 2.2 GOES satellite

The GOES satellite was launched by NASA and operated by NOAA. The flare magnitudes of events from 1992 to 2013 July have been digitized at Nobeyama Radio Heliograph Event List.[2] Regarding the events that occurred before 1992 June and after 2014 August, we refer to the GOES X-ray sensor (XRS) report[3] and the Nobeyama Radio Polarimeter Event List.[1] The XRS onboard the GOES satellite has observed the solar SXR. However, in case of GOES-16, the XRS detector contained Si photodiodes instead of ionization cells (Chamberlin et al. 2009). Thus, the spectral sensitivity of XRS varied between GOES-15 and GOES-16. This study did not consider the data after GOES-16. To remove the SWPC (Space Weather Prediction Center) scaling factors and obtain the true flux, the long band flux must be divided by 0.7 (J. Machol et al. 2022).[4]

## 3 Results

We first derive the correlation between the peaks of microwave and SXR radiation, the correlation coefficient, and fitting function assuming a power function (subsection 3.1). The total number of above M-class flares observed at the Nobeyama Radio Polarimeters and Toyokawa Observatory is 341 from 1988 to 2014, and these flares were analyzed in this study. Secondly, the correlation between the microwave and SXR peaks can be analyzed further by segmenting the type of microwave radiation. Then, we got the result showing the variations between the optically thin radiation and the optically thick radiation (subsection 3.2). Based on these results, we estimated the magnitudes and SXR light curves of the earliest solar flares (subsections 3.3) reported in the Toyokawa solar radio measurements (subsections 3.4 and 3.5).

[1] ⟨https://solar.nro.nao.ac.jp/norp/html/event/⟩.
[2] ⟨https://solar.nro.nao.ac.jp/norh/html/event/⟩.
[3] ⟨https://www.ngdc.noaa.gov/stp/space-weather/solar-data/solar-features/solar-flares/x-rays/goes/xrs/⟩.
[4] J. Machol et al. 2022, GOES X-ray Sensor (XRS) Operational Data (Version 1.5) (Stennis Space Center, NOAA) ⟨https://www.ngdc.noaa.gov/stp/satellite/goes/⟩.

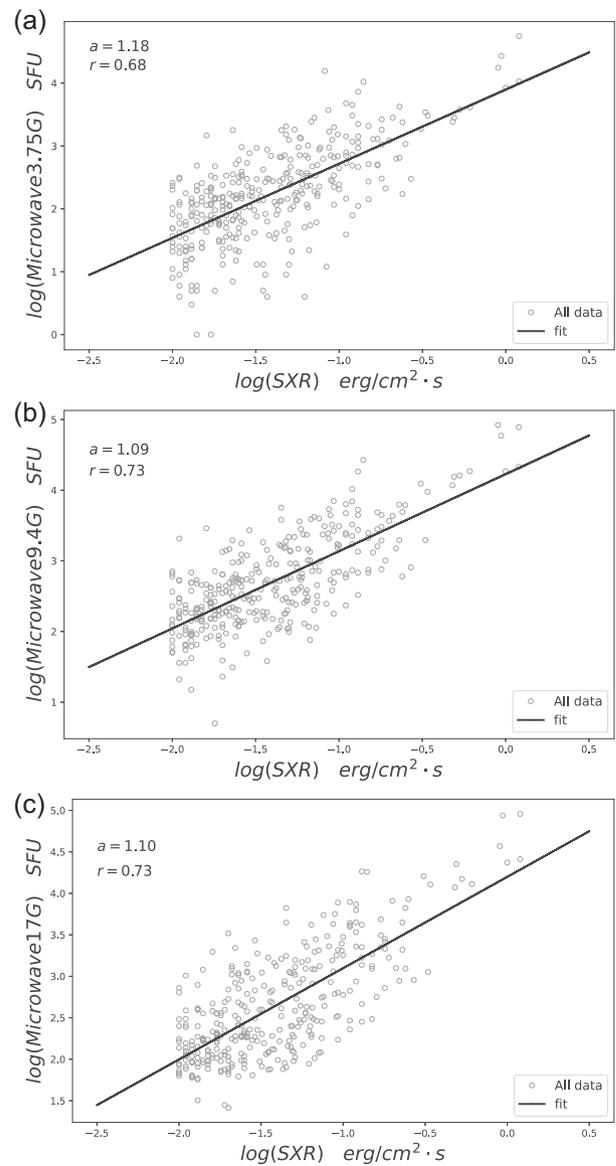

**Fig. 1.** (a) Scattering plot of the peak flux of microwave (3.75 GHz) and soft X-ray. (b) Scattering plot of the peak flux of microwave (9.4 GHz) and soft X-ray. (c) Scattering plot of the peak flux of microwave (17 GHz) and soft X-ray. In panels (a)–(c), both microwaves and soft X-rays are logarithmic values with 10 as the bottom. $r$ is the correlation coefficient and $a$ is the parameter corresponding to $a$ in equation (2).

### 3.1 Correlation between microwaves (3.75, 9.4, and 17 GHz) and SXR

The scattering plot of peak fluxes of microwave and SXR measurements is displayed in figure 1, wherein the correlation efficient $r$ is 0.68, 0.73, and 0.73 in figure 1, respectively. The fitting function is defined as

$$y = a \times x + b, \qquad (2)$$

where the logarithmic data of SXR and microwaves were put into $x$ and $y$, respectively, and the values of $a$ and $b$ were

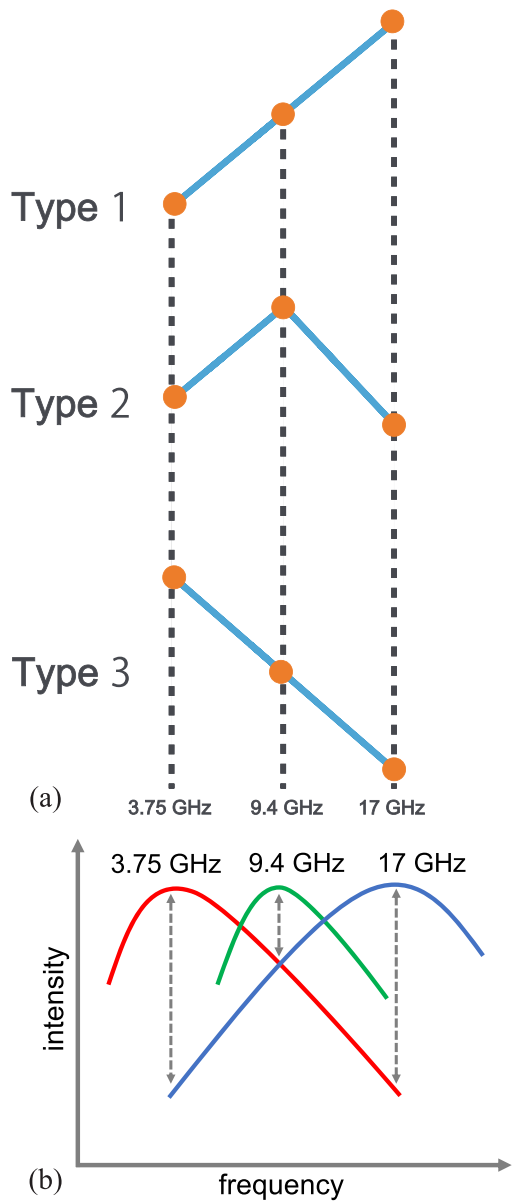

**Fig. 2.** (a) Three types divided by the microwave spectrum. For Type1, the radiation at 3.75 and 9.4 GHz is optically thick at least. For Type2, the radiation at 3.75 GHz is optically thick, 9.4 GHz is considered to be around the turnover frequency (the border between optically thick and thin in the spectrum), and 17 GHz is optically thin. For Type3, the radiation at 9.4 and 17 GHz is optically thick at least. (b) The left, middle, and right lines are examples of the spectrum of Type1, 2, and 3 flares, respectively. Around 9.4 GHz, there is not so much difference in radiation intensity if the type is different compared with those at 3.75 GHz and 17 GHz, as vertical dotted arrows indicate.

**Table 1.** Optical thickness of microwave spectrum corresponding to figure 2.

|  | 3.75 GHz | 9.4 GHz | 17 GHz |
| --- | --- | --- | --- |
| Type1 | Thick | Thick | Thick or Thin |
| Type2 | Thick | Thick or Thin | Thin |
| Type3 | Thick or Thin | Thin | Thin |

**Table 2.** Number of flare events for each type.*

| Type1 | Type2 | Type3 | GroupA | GroupB |
| --- | --- | --- | --- | --- |
| 132 | 183 | 26 | 315 | 209 |

*GroupA (Type1 + Type2) is used for optically thick radiation at 3.75 GHz. GroupB (Type2 + Type3) is used for optically thin radiation at 17 GHz.

derived from linear approximation. However, flare events cannot all be analyzed with equal integrity, because the optical thickness of the radio sources varies individually. In addition to the optical thickness, the parameters related to the microwave radiation are the magnetic field, the energy spectrum of the accelerated electrons, and the location of a flare in the solar disk. For past flares, only microwave fluxes at a few discrete frequencies are available. The only parameter that can be considered in modern observations when applied to past flares is the optical thickness obtained from the microwave spectrum. Thus, we can derive only rough spectral information, such as the turn-over frequency (boundary between optically thick and thin). Considering this situation, we focused on only optical thickness (thick or thin) in this study. Accordingly, the flare events should be appropriately analyzed in multiple groups because the observed microwave intensity depends on their optical thickness. Therefore, the data were segmented into three groups (subsection 3.2) considering the optical thickness of microwave radiation. In general, during flares, nonthermal electrons are produced, resulting in strong gyrosynchrotron radiation (Dulk 1985). Since we focused on the peak flux of the solar flares above M-class, we assume that gyrosynchrotron radiation is dominant. The microwave spectrum reflects the optical thickness of radiation. If the spectrum decreases in intensity with increasing frequency, the radiation is optically thin, whereas if the intensity increases as the frequency increases, it is optically thick.

### 3.2 Correlation between microwaves and SXR in case of considering optically depth

In figure 2a, the microwave spectrum is classified into three types of flare cases. For each type, the microwave radiation at each frequency is optically thick or thin, as shown in table 1. Here, Type1 and Type2 are both optically thick at 3.75 GHz, and Type2 and Type3 are both optically thin at 17 GHz. As they do not require distinction, GroupA (Type1 + Type2) and GroupB (Type2 + Type3) are newly added. The number of events for each type is listed in table 2. The correlation coefficients for each type are listed in table 3.

From figure 3a, the relationship between Type3 and GroupA can be explained based on the variations in optical thickness of the microwave radiation. Upon comparing the fit line of GroupA with that of Type3 based on the same strength of SXR, the microwave value of the GroupA fit line was less than that of the Type3 fit line. This is consistent with the fact that if the radiation is weaker than originally expected because of the optically thick microwave radiation, only the microwave radiation emitted from the front will be observed.

**Table 3.** Fitting parameters [$a$ and $b$ in equation (2)], correlation coefficient $r$ between the peaks of microwave and SXR, and 95% confidence interval of $r$.

|  | $a$ | $b$ | $r$ |
| --- | --- | --- | --- |
| 3.75 GHz (Type3) | 0.89 ± 0.22 | 3.99 ± 0.32 | 0.64 [0.34,0.82] |
| 3.75 GHz (GroupA) | 1.20 ± 0.07 | 3.89 ± 0.10 | 0.70 [0.63,0.75] |
| 3.75 GHz (all data) | 1.18 ± 0.07 | 3.90 ± 0.10 | 0.68 [0.62,0.73] |
| 9.4 GHz (Type1) | 1.21 ± 0.09 | 4.32 ± 0.13 | 0.77 [0.69,0.83] |
| 9.4 GHz (Type2) | 1.06 ± 0.08 | 4.24 ± 0.11 | 0.72 [0.64,0.78] |
| 9.4 GHz (Type3) | 0.69 ± 0.19 | 3.55 ± 0.28 | 0.59 [0.26,0.80] |
| 9.4 GHz (all data) | 1.09 ± 0.06 | 4.23 ± 0.08 | 0.73 [0.68,0.78] |
| 17 GHz (Type1) | 1.11 ± 0.08 | 4.39 ± 0.12 | 0.77 [0.69,0.83] |
| 17 GHz (GroupB) | 1.01 ± 0.07 | 3.96 ± 0.11 | 0.71 [0.63,0.77] |
| 17 GHz (all data) | 1.10 ± 0.06 | 4.20 ± 0.08 | 0.73 [0.68,0.78] |

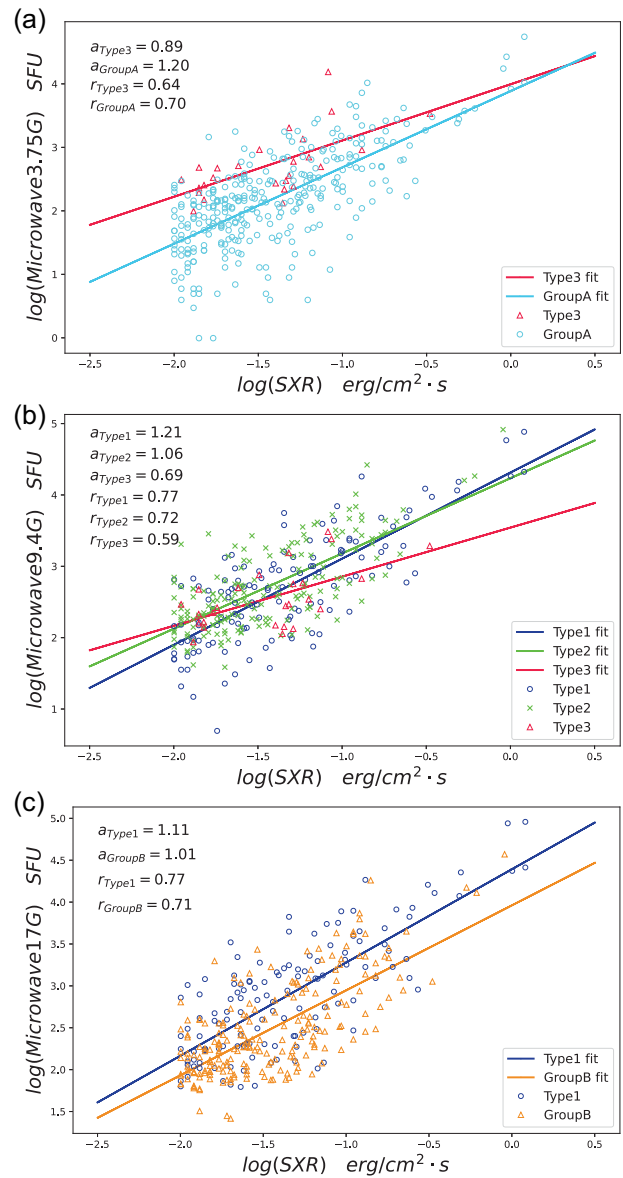

**Fig. 3.** Scatter plots similar to figure 1, if types are created. (a)–(c) Scattering plot of peak flux of microwave (3.75–17 GHz) and soft X-ray (SXR). In panels (a)–(c), both microwaves and SXR are logarithmic values with 10 as the bottom, and $r$ and $a$ denote the fitting results, similar to figure 1.

In figure 3b, we consider that the fit lines obtained for Type1, Type2, and Type3 do not differ between these types. In Silva, Wang, and Gary (2000), the most common turnover frequency was approximately 10 GHz. For several events considered in this study, 9.4 GHz is proximate to the turnover frequency. Therefore, around 9.4 GHz, the three types are considered to be less affected by optical thickness compared with the situations at 3.75 GHz and 17 GHz as shown in figure 2b. The classification by microwave spectral shape does not reflect the effect of optical thickness, and no difference can be observed between the Type1, Type2, and Type3 fit lines for 9.4 GHz.

The observations in figure 3c cannot be explained by the variations in optical thickness. As the 17 GHz microwave radiation was largely optically thin, we conceived that Type1 emissions were optically thin and Type1 and GroupB emissions cannot be separated. To determine the optical thickness of 17 GHz emissions accurately, further extensive data are required at frequencies higher than 17 GHz. As an interpretation of figure 3c, the flares of Type1 are considered to be non-thermal and rich compared with those of GroupB. The difference between Type1 and GroupB in 17 GHz may be affected by the tendency of Type1 flares characterized by their non-thermal richness than the optical thickness. The high turnover frequency of Type1 indicates a dense ambient plasma and strong magnetic field. While the relationship between non-thermal rich flares and high turnover frequency is not clear and should be studied, it is not a topic of this study.

### 3.3 Overview of great solar flares in 1956

This study estimated the flare magnitude of two solar flares on 1956 February 14 and 23 from Toyokawa solar radio measurements. For the flare that occurred on 1956 February 23, in Tanaka and Kakinuma (1956) the initiation time was 03:34:00 UT, end time was 03:49:30 UT, and peak time was 03:35:25 UT. This flare emitted violent radio radiation; the maximum intensity was 225 and 141 times as large as that of the quiet-sun radio emission at 3.75 and 9.4 GHz, respectively. The flare observed on 1956 February 14 is described in Tanaka et al. (1956). For 3.75 and 9.4 GHz radiation, the peak values were approximately 9400 and 2700 (SFU = $10^{-22}$ W m$^{-2}$ Hz$^{-1}$), respectively.

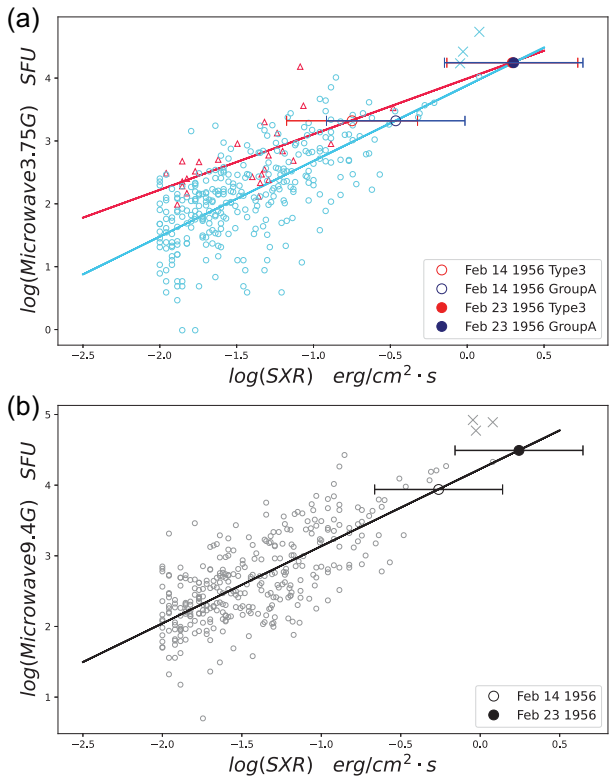

**Fig. 4.** Estimated flare classes on 1956 February 14 and 23. (a) The 1956 flares are located at the red and blue circle points corresponding to Type3 and GroupA. (b) The 1956 flares are located at the black circle points. In panels (a) and (b), the three cross marks are the "microwave rich events" discussed in subsection 4.3.

**Table 4.** Estimated value (EV) of the flare class for 3.75 GHz (Type3), 3.75 GHz (GroupA), and 9.4 GHz.[*]

|  | 3.75 GHz (Type3) | 3.75 GHz (GroupA) | 9.4 GHz |
|---|---|---|---|
| | — February 23 — | | |
| L L (1$\sigma$) | X6.5 | X8.5 | X7.5 |
| E V | X20 | X20 | X18 |
| U L (1$\sigma$) | X59 | X48 | X41 |
| | — February 14 — | | |
| L L (1$\sigma$) | M5.9 | X1.4 | X2.3 |
| E V | X1.8 | X3.4 | X5.5 |
| U L (1$\sigma$) | X5.4 | X8.1 | X13 |

[*]LL (lower limit) and UL (upper limit) indicate ranges of 1$\sigma$. Each EV for these three patterns corresponds to points of 1956 flares in figure 4.

These flares included fundamental solar flares among the earliest solar radio bursts. They were missed from the direct measurements of the GOES satellites because GOES started operations only after the mid-1970s. However, the microwave observation was under operation at frequencies of 3.75 and 9.4 GHz. The light curves for the flares observed on 1956 February 23 at 3.75 and 9.4 GHz are plotted in Tanaka et al. (1956), which also presents the light curves for the flares that occurred on 1956 February 14.

### 3.4 Applications to the early solar flares: case studies

The peak fluxes observed at 3.75 and 9.4 GHz were used to predict the class of the earliest solar radio bursts at Toyokawa Observatory. Herein, we traced and digitized the light curves of the earliest solar radio observations reported by Toyokawa Observatory in Tanaka and Kakinuma (1956). For 3.75 GHz, we used the fitting function obtained in figure 3a to estimate the two types of flare class. For 9.4 GHz, we decided to use the fitting function obtained in figure 1b as we did not consider it necessary to separate the flares into several types. Consequently, as listed in table 4, the flare class on 1956 February 23 was determined as approximately X20. For this flare event, the estimated flare class was consistent with Cliver et al. (2020). However, the large variances in fitting functions were caused by variations in estimating the flare class of a factor of 3. Moreover, we digitized from a observational record on 1956 February 14 in Toyokawa Observatory and estimated the flare class as well as the flare on 1956 February 23. The flare class was approximately X2–X6 (table 4). Note that these flare classes need to be divided by 0.7 if we apply the correction reported by J. Machol et al. (2022).[4]

### 3.5 Reproduction of the SXR light curve on 1956 February 14 and 23 for earliest solar radio bursts at Toyokawa observatory

The light curve of SXR can be estimated using equation (1) if the light curve of the non-thermal microwave radiation is known. First, we digitized the light curve of the microwaves reported by Toyokawa Observatory (Tanaka & Kakinuma 1956) from the figure and produced the data at 3.75 and 9.4 GHz, as presented in figures 5a–5b and figures 6a–6b. The baseline before the burst was determined and the variation from the baseline was observed as $\Phi_{\text{Micro}}$. Secondly, we decided on the shape of the SXR light curve following equation (1). Finally, the estimated flare class was normalized using the fitting function at each frequency and the SXR light curve was quantitatively determined.

To reproduce the SXR light curves from the nonthermal radiation light curves according to the Neupert effect, two things must be considered: setting the peak time of the SXR light curve and the flare decay period as well as the rise period cannot be reproduced using the Neupert effect. As reported by Veronig et al. (2002a), the termination period of the nonthermal radiation in solar flares coincided with the peak time of the thermal radiation. Therefore, this study assumed that the end time of microwave radiation coincides with the peak time of SXR emission. In addition, we assumed that the SXR light curve starts to diminish in case the cooling rate $\tau_c$ exceeds the rising rate of the thermal radiation light curve $\tau_r$ owing to the nonthermal radiation. Moreover, the end time of the microwave light curve is when $\tau_c \simeq \tau_r$ holds, and $\tau_c$ and $\tau_r$ can be defined by the following equations:

$$\tau_r = \frac{\partial F_{SXR}}{\partial t}, \quad \tau_c = \frac{F_{peak}}{2T_{decay}}, \tag{3}$$

where $F_{SXR}$, $F_{peak}$, and $T_{decay}$ denotes the flux of SXR, the peak flux of SXR, and the decay duration of the solar flares on the GOES satellites, respectively. The cooling rate was obtained by considering that the median value of the decay period $T_{decay}$ was 14 min in X flares in Veronig et al. (2002b) and the decay period of a GOES flare was defined as the time required to decay to a value that is intermediate between the peak flux and background. The effect of radiative cooling was considered to be significant (Antiochos & Sturrock 1978; Culhane et al. 1994). The rate of increase was determined at each instant by the SXR curve plotted by the end time of the nonthermal microwave emission and the estimated flare class. Subsequently, with respect to the flare decay period, we plotted a dotted line in figures 6c–6d and 7c–7d under the assuming constant cooling rate. Although the radiative cooling rate can be obtained assuming the temperature, density, and length of the flare loop (Aschwanden 2007), it is beyond the fundamental argument of this study, and thus is neglected. This result is the first study that describes the SXR light curve before initiating the observation of SXR. We reconstructed the SXR light curves using the non-thermal microwave emission at 3.75 and 9.4 GHz. They show similar variation. This fact increases confidence while it is not possible to determine which one is more accurate. As described in the introduction, the SXR light curves are associated with the length of the loop and the duration of magnetic reconnection (Reep & Toriumi 2017). In this paper, we only reconstructed the SXR light curve of the flares in 1956. The physical consideration can be developed for these flares in the future.

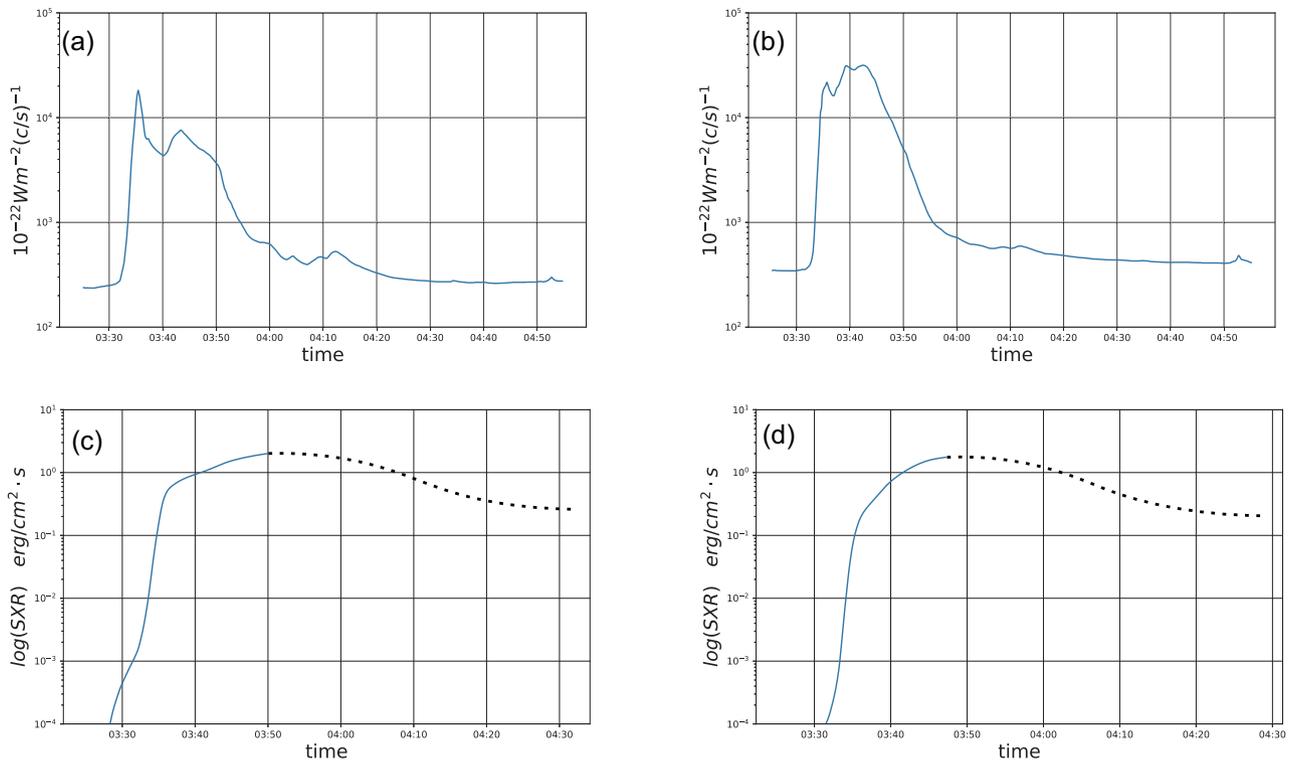

**Fig. 5.** Digitized microwave data and estimated SXR light curve on 1956 February 23. (a) Digitized microwave light curve at 3.75 GHz. (b) Digitized microwave light curve at 9.4 GHz. (c) Estimated soft X-ray light curve derived from microwave data at 3.75 GHz assuming optically thick microwave radiation at 3.75 GHz. (d) Estimated soft X-ray light curve derived from microwave data at 9.4 GHz. In panels (a)–(d), time is in Coordinated Universal Time (UTC).

## 4 Discussion

### 4.1 Influence of location on the occurrence of flare

As described in subsection 3.1, microwave radiation is related to various parameters, not just optical thickness.

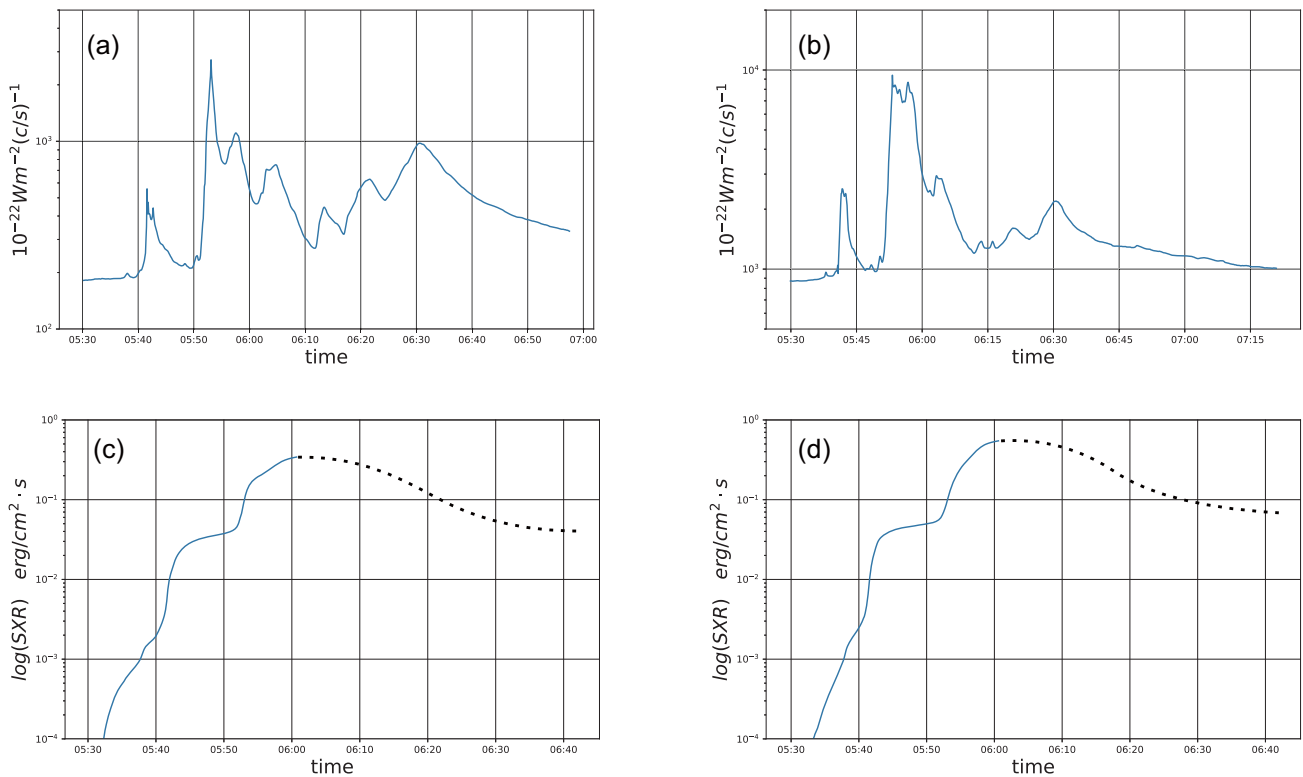

**Fig. 6.** Digitized microwave data and estimated soft X-ray light curve on 1956 February 14. (a) Digitized microwave light curve at 3.75 GHz. (b) Digitized microwave light curve at 9.4 GHz. (c) Estimated soft X-ray light curve derived from microwave data at 3.75 GHz assuming optically thick microwave radiation at 3.75 GHz. (d) Estimated soft X-ray light curve derived from microwave data at 9.4 GHz. In panels (a)–(d), time is in UTC.

In this study, we did not regard the location of the flares when selecting the flare events. As for the influence of the location, it wouldn't affect the intensity of the microwaves when analyzing the flares without classifying them by the parameters related to the microwave radiation (Silva & Valente 2002; Kawate et al. 2011). Overall, we can neglect the influence of the position by estimating the flare from the microwave data at 3.75 and 9.4 GHz.

### 4.2 Accuracy of SXR light curve and impact of SXR background flux

We cannot reproduce the decay phase of the SXR light curve using the Neupert effect because the effect continues to integrate the nonthermal light curve until the end time of the flare. The starting and ending times of the reproduced SXR light curve depend on the time and duration of the nonthermal radiation, which complicates the precise identification of the time variations in the SXR light curve.

The microwave peak flux data in the NoRP Event List represents the peak flux after background subtraction. However, the GOES peak flux was not subtracted by the background level. In particular, small flares (-B class) are influenced by the SXR background flux (Wheatland 2010). Moreover, as the SXR background flux is large in the case of a high solar cycle, the small flares cannot be easily detected and the background level depends on the solar cycle (Wheatland & Litvinenko 2002). Thus, the SXR peak flux subtracted data can be efficiently used. In accordance with Hudson (2021), this study considered solar flares above M-class to minimize the influence from the background variations.

### 4.3 Microwave-rich solar flare events

In subsections 3.1 and 3.2, the correlation between peak fluxes of microwave and SXR is statistically indicated. There are several events with a large flare class which are far from the correlation line. Some of them might be considered as microwave-rich events. Previous studies reported that the flares exhibit a higher proportion of nonthermal radiation and less thermal radiation (Fleishman et al. 2011; Masuda et al. 2013). Three large flare events with rich microwave radiation are depicted in figure 4, one of which is the X12-class flare on 1991 June 6. However, the light curve of the GOES SXR flux (1–8 Å) on this flare was saturated (H. Hudson 2022 private communication). Considering the saturation, the actual flare class will be higher and will correspond well with the fitting line. The other two cases include the flares on 1991 March 22 and 1992 November 2, with classes of X9.4 and X9, respectively. As they estimated a long duration with a large flare class, we believe that several nonthermal electrons were trapped by the large magnetic loops and they emitted microwave for a long duration (Masuda et al. 2013).

## 5 Conclusion

This study examined the statistical trends using microwave peak fluxes with multiple NoRP frequencies and GOES SXR peak fluxes observed between 1988 and 2014. The results revealed that the correlation coefficient was 0.7 for all microwave frequencies. We further classified the flares by considering the optical thickness of the microwave radiation. The effect of optical thickness was observed at 3.75 GHz, whereas at 9.4 GHz the flares were not affected by the optical thickness, probably near the turnover frequency. Based on the known relationship between SXR and three microwave frequencies, the scales of the flares on 1956 February 14 and 1956 February 23 were estimated as X1.8–X5.5 (ranging between M5.9–X13) and X18–X20 (ranging between X6.5–X59), based on the fitting functions derived using the microwave peak flux data at instant time (3.75 and 9.4 GHz). Their variances were large and varied by a factor of 3 in $1\sigma$. Using the Neupert effect, the shape of the SXR light curves was estimated from the microwave light curve, and the light curve of the SXR was quantitatively drawn by normalizing it by the estimated flare class.

## Acknowledgments

This work is partly supported by JSPS KAKENHI, grant JP18H01253. Hisashi Hayakawa (HH) was financially supported in part by JSPS Grants-in-Aid JP20K22367, JP20K20918, JP20H05643, and JP21K13957; the JSPS Overseas Challenge Program for Young Researchers; the ISEE director's leadership fund for FY2021; the Young Leader Cultivation (YLC) program; the young researcher units for the advancement of new and undeveloped fields of the Institute for Advanced Research (Nagoya University) under the Program for Promoting the Enhancement of Research Universities; and Tokai Pathways to Global Excellence (Nagoya University) of the Strategic Professional Development Program for Young Researchers (MEXT). HH was benefitted from discussions in the ISSI International Teams ("Solar Extreme Events: Setting Up a Paradigm" and "Modeling Space Weather and Total Solar Irradiance Over The Past Century"). HH thanks Hugh Hudson, Michael Wheatland, and Edward Cliver for letting him have fruitful discussions.

## References

Antiochos, S., & Sturrock, P. 1978, ApJ, 220, 1137
Aschwanden, M. J. 2007, ApJ, 661, 1242
Aschwanden, M. J. 2012, Space Sci. Rev., 171, 3
Chamberlin, P. C., Woods, T. N., Eparvier, F. G., & Jones, A. R. 2009, SPIE Proc., 7438, 743802
Cliver, E. W., & Dietrich, W. F. 2013, J. Space Weather Space Clim., 3, A31
Cliver, E. W., Hayakawa, H., Love, J. J., & Neidig, D. F. 2020, ApJ, 903, 41
Cliver, E. W., Schrijver, C. J., Shibata, K., & Usoskin, I. G. 2022, Living Rev. Sol. Phys., 19, 2
Culhane, J. L., et al. 1994, Sol. Phys., 153, 307
Curto, J. J., Castell, J., & Del Moral, F. 2016, J. Space Weather Space Clim., 6, A23
Dennis, B. R., & Zarro, D. M. 1993, Sol. Phys., 146, 177
Doherty, P., Coster, A. J., & Murtagh, W. 2004, GPS Solutions, 8, 267
Dulk, G. A. 1985, ARA&A, 23, 169
Elvidge, S., & Angling, M. J. 2018, Space Weather, 16, 417
Fleishman, G. D., Kontar, E. P., Nita, G. M., & Gary, D. E. 2011, ApJ, 731, L19
Fletcher, L., et al. 2011, Space Sci. Rev., 159, 19
Gonzalez, W. D., Joselyn, J. A., Kamide, Y., Kroehl, H. W., Rostoker, G., Tsurutani, B. T., & Vasyliunas, V. M. 1994, J. Geophys. Res., 99, 5771
Hayakawa, H., et al. 2019, Space Weather, 17, 1553
Hayakawa, H., et al. 2022a, MNRAS, 517, 1709
Hayakawa, H., Nevanlinna, H., Blake, S. P., Ebihara, Y., Bhaskar, A. T., & Miyoshi, Y. 2022b, ApJ, 928, 32
Hudson, H. S. 1991, BAAS, 23, 1064
Hudson, H. S. 2021, ARA&A, 59, 445
Hudson, H. S., & Ohki, K. 1972, Sol. Phys., 23, 155
Ivanov, K. G., & Kharshiladze, A. F. 2007, Geomagn. Aeron., 47, 787
Kawate, T., Asai, A., & Ichimoto, K. 2011, PASJ, 63, 1251
Masuda, S., Shimojo, M., Kawate, T., Ishikawa, S., & Ohno, M. 2013, PASJ, 65, S1
Nakajima, H., Sekiguchi, H., Sawa, M., Kai, K., & Kawashima, S. 1985, PASJ, 37, 163
Neupert, W. M. 1968, ApJ, 153, 59
Notsuki, M., Hatanaka, T., & Unno, W. 1956, PASJ, 8, 52
Reep, J. W., & Toriumi, S. 2017, ApJ, 851, 4
Rishbeth, H., Shea, M. A., & Smart, D. F. 2009, Adv. Space Res., 44, 1096
Shea, M. A., & Smart, D. F. 2012, Space Sci. Rev., 171, 161
Shimojo, M., & Iwai, K. 2023, Geoscience Data J., 10, 114
Silva, A. V. R., & Valente, M. M. 2002, Sol. Phys., 206, 177
Silva, A. V. R., Wang, H., & Gary, D. E. 2000, ApJ, 545, 1116
Tanaka, H., & Kakinuma, T. 1956, Proc. Res. Inst. Atmos., Nagoya University, 4, 74
Tanaka, H., & Kakinuma, T. 1957, Proc. Res. Inst. Atmos., Nagoya University, 4, 60
Tanaka, H., Kakinuma, T., Jindo, H., Takayanagi, T., & Torii, C. 1956, Bull. Res. Inst. Atmos., Nagoya University, 6, 61
Tsiftsi, T., & Luz, V. D. l. 2018, Space Weather, 16, 1984
Usoskin, I. G. 2017, Living Rev. Sol. Phys., 14, 3
Usoskin, I. G., Koldobskiy, S. A., Kovaltsov, G. A., Rozanov, E. V., Sukhodolov, T. V., Mishev, A. L., & Mironova, I. A. 2020, JGR: Space Phys., 125, e27921
Veronig, A., Temmer, M., Hanslmeier, A., Otruba, W., & Messerotti, M. 2002b, A&A, 382, 1070

Veronig, A., Vršnak, B., Temmer, M., & Hanslmeier, A. 2002a, Sol. Phys., 208, 297

Veronig, A. M., Brown, J. C., Dennis, B. R., Schwartz, R. A., Sui, L., & Tolbert, A. K. 2005, ApJ, 621, 482

Wheatland, M. S. 2010, ApJ, 710, 1324

Wheatland, M. S., & Litvinenko, Y. E. 2002, Sol. Phys., 211, 255

Yashiro, S., & Gopalswamy, N. 2009, in IAU Symp. 257, Universal Heliophysical Processes, ed. N. Gopalswamy & D. F. Webb (Cambridge: Cambridge University Press), 233